\newcommand{\be}{\begin{equation}}
\newcommand{\ee}{\end{equation}}
\newcommand{\bea}{\begin{eqnarray}}
\newcommand{\eea}{\end{eqnarray}}
\newcommand{\beas}{\begin{eqnarray*}}
\newcommand{\eeas}{\end{eqnarray*}}

\newcommand{\para}{||}
\newcommand{\sss}{{s}}
\newcommand{\sy}{{\mbox{{\bf \scriptsize{Y}}}}}

\newcommand{\sye}{{\mbox{{\bf \tiny{Y}}}}}
\newcommand{\ext}{{\mbox{\tiny{ext}}}}

\newcommand{\slsh}[1]{{\not \! #1}}
\newcommand{\slshh}[1]{{\not \!\! #1}}

\documentclass[twocolumn,aps,showpacs,nofootinbib,epsfig]{revtex4}
\usepackage{graphics}
\usepackage{epsfig}
\begin{document}

\title{Effective potential at finite temperature in a constant
hypermagnetic field: Ring diagrams in the Standard Model}

\author{Angel Sanchez$^\dagger$, 
        Alejandro Ayala$^\dagger$ 
        and Gabriella Piccinelli$^\ddagger$ 
        }    

\affiliation{$^\dagger$Instituto de Ciencias Nucleares, Universidad
        Nacional Aut\'onoma de M\'exico, Apartado Postal 70-543, M\'exico
        Distrito Federal 04510, M\'exico.\\
        $^\ddagger$Centro Tecnol\'ogico, FES Arag\'on,
        Universidad Nacional Aut\'onoma de M\'exico,
        Avenida Rancho Seco S/N, Bosques de Arag\'on,
        Nezahualc\'oyotl, Estado de 
        M\'exico 57130, M\'exico.}

\begin{abstract}

We study the symmetry breaking phenomenon in the standard model during the
electroweak phase transition in the presence of a constant
hypermagnetic field. We compute the finite temperature effective potential up
to the contribution of ring diagrams in the weak field, high temperature limit
and show that under these conditions, the phase transition becomes stronger
first order. 

\end{abstract}


\maketitle

\section{Introduction}\label{I}

The problem of baryogenesis is still one of the outstanding open questions in 
cosmology, despite the large amount of work devoted to find a viable
explanation. The conditions for developing a baryon asymmetry in an initially 
symmetric universe were established by Sakharov in 1967~\cite{Sakharov}
and the search for a scenario to encompass them continues. These three
well-known conditions are: (1) existence of 
interactions that violate baryon number; (2) C and CP violation and
(3) departure from thermal equilibrium.    

The Standard Model (SM) of electroweak interactions meets all these
requirements, provided the electroweak phase transition (EWPT) be
first order, since, at that stage of the universe evolution, this is
the only possible source of departure from thermal equilibrium. 
Nonetheless, it is well known that neither the amount of CP violation within
the minimal SM~\cite{Gavela}, nor the strength of the EWPT are enough to
generate a sizable baryon number~\cite{Kajantie1}. 

During the EWPT, the Higgs field vacuum expectation value changes from zero
(false vacuum) in the symmetric phase, to a finite value $\langle v\rangle$
(true vacuum) in the broken phase. This evolution is determined by the finite
temperature effective potential of the theory, that develops a barrier between
the two minima if the phase transition is first order. The temperature at
which the two minima are degenerate is called the critical temperature $T_c$
and this instant is considered the beginning of the phase transition. From
there on, the transition is accomplished through the nucleation, expansion and
percolation of true vacuum bubbles in the background of false vacuum, leading
to a departure from equilibrium conditions. 

The existence of baryon number violation is realized in the SM by
means of its vacuum structure through {\it sphaleron} mediated processes.
The sphaleron~\cite{Klinkhamer} is a static and unstable solution of the
field equations of non-Abelian gauge theories, corresponding to the top of the
energy barrier between topologically distinct vacua, where the minima
correspond to configurations with zero gauge field energy but
different baryon number. Transitions are associated to baryon number
$n_{B}-n_{\bar{B}}$ violation and can either induce or wash
out a baryon asymmetry.  

The above property of sphalerons makes the preservation of a given baryon
asymmetry one of the most difficult conditions to meet during the baryogenesis
process in the majority of the proposed scenarios. This requires that the
baryon violating transition between different topological vacua is suppressed
in the broken phase, when the universe returns to thermal equilibrium. 
In other words, the sphaleron transitions must be slow
compared to the expansion rate of the universe and this in turn
translates into the condition $\langle v\rangle/T_c\geq 1.0 -
1.5$~\cite{Kajantie1}. Although the above is an
estimate emerging from approximate calculations, it is nowadays widely
accepted and has proven to be a rather difficult condition to meet.

Although there have been several attempts to link the baryogenesis process to
the EWPT~\cite{reviewsEWPT, Petropoulos} in general these all share the
characteristic that 
the Sakharov conditions are only partially met. Here we want to further
explore the possibility to embed the baryogenesis process in the EWPT
scenario, including the effect of an extra ingredient: the possible presence
of primordial magnetic fields. Before the EWPT, magnetic fields couple to
matter through the particle's hypercharge and thus properly receive the name
of {\it hypermagnetic} fields.

Magnetic fields seem to pervade the entire universe and their
generation may be either primordial or associated to the process of
structure formation. They have been observed in galaxies, clusters,
intracluster medium and high redshift objects~\cite{31}. In order to
distinguish between primordial and protogalactic fields, it is useful to
search for their imprint on the cosmic microwave background radiation
(CMBR). A homogeneous magnetic field would 
give rise to a dipole anisotropy in the background radiation, on this basis,
Cosmic Background Explorer (COBE) results
give an upper bound on the  present equivalent field strength of $B_0\lesssim
10^{-9}$ G~\cite{Barrow}. An upper limit to the field strength of $B_0\lesssim
4.7\times 10^{-9}$ G at the present scale of 1 Mpc is obtained by an analysis
that includes small scale CMBR anisotropies from Wilkinson Microwave
Anisotropy Probe (WMAP), Cosmic Background Imager (CBI) and Arc Minute
Cosmology Bolometer Array Receiver (ACBAR)~\cite{Yamazaki}. On the other hand,
tangled random fields on small scales could reach up to $\sim 10^{-6}$
G~\cite{Barrow2}. Nucleosynthesis also imposes limits on
primordial magnetic fields since they have an influence on both the universe
expansion rate and the electron quantum statistics. The observed helium
abundance implies $B \lesssim 10^{12}$ G at scales greater than 10 cm at the
end of nucleosynthesis~\cite{Grasso}.

Although at present there is no conclusive evidence about the origin
of magnetic fields, their existence prior to the EWPT cannot certainly
be ruled out making it important to investigate their effect on the
baryogenesis process~\cite{elreview}. In fact, it has been
shown that these fields provide mechanisms to affect
all the Sakharov conditions: In the presence of a magnetic field, the
phase transition becomes stronger first order in analogy to the case of a type
I superconductor, where the Meissner effect brings the phase transition from
second to first order~\cite{Giovannini,Elmfors,Kajantie}; it has also been
shown that extra CP violation is 
obtained from the segregation of axial charge during the reflection and
transmission of fermions through the vacuum bubbles due to the chiral nature of
their coupling to the hypermagnetic field~\cite{Pallares}; finally, regarding
the sphaleron transitions the presence of a magnetic field works against the
preservation of a baryon asymmetry due to the coupling between the sphaleron
dipole moment and the magnetic field that lowers the energy barrier between
topologically distinct minima~\cite{Dario}.

The effect of magnetic fields on the EWPT has been analytically studied both
classically ~\cite{Giovannini} and to one-loop order~\cite{Elmfors}, as well
as by means of lattice simulations~\cite{Kajantie}. These calculations all
agree that the strength of the phase transition is enhanced by the presence of
hypermagnetic fields, although the ratio $\langle v\rangle/T_c$ does not reach
the desired value, for a large Higgs boson mass. On the other hand, other
analytical approaches where the SM finite temperature effective potential is
studied for the case of strong magnetic fields~\cite{Skalozub1,Skalozub2},
reach the conclusion that these fields inhibit the first order phase
transition and attribute the result to the contribution of light fermion
masses which are generally neglected in other computations.

In this work we concentrate on studying the relation between the presence of a
large scale magnetic field and the dynamics of the EWPT by computing the SM
finite temperature effective potential in a constant hypermagnetic
field up to the contribution of ring diagrams, that have been shown to be
crucial for the description of the long wavelength properties of the
theory~\cite{Carrington}. We carry out a systematic calculation of each SM
sector showing that the major contribution producing an enhancement of the
EWPT comes from the Higgs and gauge boson sectors and that fermions do not act
against this behavior. Working in the limit where the magnetic field is weak,
we find an enhanced value of the ratio $\langle v\rangle/T_c$.

The paper is organized as follows: In Sec.~\ref{II} we lay down
the formalism to include weak magnetic fields in the computation of
hypercharged particle propagators. In Sec.~\ref{III} we write down the SM
using the degrees of freedom in the symmetric phase. In Sec.~\ref{IV}, we
work with these degrees of freedom to compute particle self energies that
are used in Sec.~\ref{V} to compute the SM effective potential
up to the contributions of ring diagrams. In Sec.~\ref{VI} we study this
effective potential as a function of the Higgs vacuum expectation value and
show that the order of the EWPT becomes stronger first order in the presence of
the hypermagnetic field. Finally, we conclude and discuss our results in
Sec.~\ref{VII} and leave for the appendix the computation of some
intermediate results of the analysis.

\section{Charged particle propagators in the presence of a Hypermagnetic
Field}\label{II} 

We work with the degrees of freedom of the SM in the  symmetric phase, where
the external (hyper)magnetic field belongs to the $U(1)_Y$ group.
To include the effect of the external field, we use Schwinger's proper time 
method~\cite{Schwinger}. In the symmetric phase, we have only two kinds of
hypercharged particles that couple to the external field namely, scalars
and fermions, whose propagators are  
\bea
    D_{H}(x,x') = \phi(x,x') \int \frac{d^4 K}{(2 \pi)^4} e^{-i k \cdot
                 (x-x')} D_{H}(k),
\label{scalprop}
\eea
\bea
    S_{H}(x,x') = \phi(x,x') \int \frac{d^4 K}{(2 \pi)^4} e^{-i k \cdot
                 (x-x')} S_{H} (k),
\label{ferprop}
\eea
respectively. The phase factor $\phi(x,x')$, that breaks translation
invariance, is given by 
\bea
\phi (x,x') &\equiv& e^{i \sye \int^x_{x'} d\xi^\mu \left[B^{\ext}_\mu +\frac{1}{2}
F_{\mu\nu} (\xi-x')^{\nu}\right]},
\label{phase}
\eea
where the vector potential $B^{\ext}_\mu=\frac{H}{2}(0,y,-x,0)$ gives rise
to a  constant hypermagnetic field of strength $H$ along the $\hat{z}$ axis
and $F^{\ext}_{\mu\nu}=\partial_\mu B^{\ext}_\nu-\partial_\nu B^{\ext}_\mu$ is
the external field strength tensor.

The momentum dependent functions $D_{H} (k)$ and $S_{H} (k)$ are given
by 
\bea
   i D_{H} (k)&=&\int_0^\infty \frac{ds}{\cos{ \sy H s}} 
              \nonumber \\
             &\times& 
            \exp\left\{ i s (k_{\para}^2-k_{\bot}^2
           \frac{\tan{\sy H \sss}}{\sy Hs}-m^2 
          +i \epsilon)\right\},
\label{scalpropmom}
\eea
\bea
   i S_{H} (k)&=&\int_0^\infty \frac{ds}{\cos{ \sy H s}} 
               \nonumber \\
             &\times& 
            \exp\left\{ i s (k_{\para}^2-k_{\bot}^2
           \frac{\tan{\sy H \sss}}{\sy Hs}-m^2 
          +i \epsilon)\right\}
            \nonumber \\
           &\times& \left[  (m_f-\slsh{k}_{\para})e^{i \sye H s \sigma_3}
           -\frac{\slsh{k_\bot}}{\cos{\sy H s}}\right],
\label{ferpropmom}
\eea
where $\mbox{\bf \scriptsize{Y}}$ is the particle's hypercharge and we use the
notation $k_{\para}^2=k_0^2-k_3^2$ and $k_\perp^2=k_1^2+k_2^2$. 

Since the gauge bosons do not couple to the external field their propagator is
given by 
\bea
   iD^{\mu\nu}_{ab}(k)=-i \left\{
                 \frac{g^{\mu\nu}-(1-\xi)\frac{k^{\mu}k^{\nu}}
                 {k^2-\xi m_G^2}}{k^2-m_G^2+i\epsilon}\right\}_{ab}
\eea
where $a,b=1,2,3,4$ and the first three values correspond to the $SU(2)_L$
fields  and the fourth to the $U(1)_Y$ field. Notice that the matrix $m_G^2$
is not diagonal in the basis of the weak-interacting fields in the symmetric
phase.  

It has been show that, by deforming the contour of integration,
Eqs.~(\ref{scalpropmom}) and~(\ref{ferpropmom}) can be written
as~\cite{nosotros,Tzuu}   
\bea
   iD_{H}(k)=2i\sum_{l=0}^{\infty}\frac{(-1)^lL_l(\frac{2k_\perp^2}{\sye
       H}) 
   \exp\left\{-\frac{k^2_\perp}{\sye H}\right\}}{k^2_{\para}-(2l+1)\sy
   H-m^2+i\epsilon}, 
\label{scalpropsum}
\eea
\bea
   iS_{H}(k)= i \sum^\infty_{l=0} 
           \frac{d_l(\frac{k_\perp^2}{\sye H})D + 
           d'_l(\frac{k_\perp^2}{\sye H}) \bar D}{k^2_{\para}-2
           l\sy H-m_f^2
           + i\epsilon} + \frac{\slsh{k_{\bot}}}{k^2_\perp}, 
\label{ferpropsum} 
\eea 
where
$d_l(\alpha)\equiv (-1)^n e^{-\alpha}
L^{-1}_l(2\alpha)$, $d'_n=\partial d_n/\partial \alpha$,
\bea
D &=& (m_f+\slsh{k_{\para}})+ \slsh{k_{\perp}} \frac{m_f^2-k^2_{\para}}{
{k^2_{\perp}}},\nonumber \\
\bar D &=& \gamma_5 \slsh{u}\slsh{b}(m_f + \slsh{k_{\para}}),
\label{DDe}
\eea
$L_l$, $L_l^m$ are  Laguerre and Associated Laguerre polynomials, respectively,
and $u^\mu$, $b^\mu$ are four-vectors describing the plasma rest frame and
the direction of the hypermagnetic field, respectively.

In order to set the appropriate hierarchy of energy scales we resort to
qualitative cosmological bounds on the possible strength of primordial
magnetic fields during the EWPT. CMBR sets stringent bounds on large scale
primordial fields but no so much stringent when the fields are tangled. This
dependence on the scale makes it difficult to extrapolate these bounds down to
the EWPT epoch. We use instead the requirement that the magnetic energy
density $\rho_{mag} \sim B^2$ should be smaller than the overall radiation
energy density $\rho_{rad} \sim T^4$ at nucleosynthesis, in order to preserve 
the estimated abundances of light elements. With this, one obtains the simple
bound $B\lesssim T^2$~\cite{Maartens}. 

On the other hand stability conditions against the formation of $W$-condensate 
indicate~\cite{Ambjorn} that the field strength is also weak compared to the
square of the $W$ mass, $m_W^2$. Notice however that when thermal corrections
are taken into account, this bound could be avoided~\cite{Skalozub1,
Skalozub2}.  

We work explicitly with the assumption that the  hierarchy of scales 
\bea
   \sy H \ll m^2 \ll T^2,
\label{hierarchy}
\eea  
is obeyed, were we consider $m$ as a generic mass of the problem at the
electroweak scale. 

We can thus perform a weak field expansion in
Eqs.~(\ref{scalpropsum}) and~(\ref{ferpropsum}) which allows to carry out the
summation over Landau levels to write the scalar and fermion propagators as 
power series in $\sy H$, that up to order $(\sy H)^2$ read
as~\cite{nosotros,Tzuu}
\bea
   D(k)_{H}=\frac{1}{k^2-m^2}\left( 1-\frac{(\sy H)^2}{(k^2-m^2)^2}-
             \frac{2(\sy H)^2 k_\perp^2}{(k^2-m^2)^3}\right),\nonumber\\
\label{scalpropweak}
\eea
and
\bea
   {S(k)_{H}}= \frac{\slsh{k}+m_f}{\slsh{k}^2-m_f^2}+
       \frac{\gamma_5
       \slsh{u}\slsh{b}(k_{\para}+m_f)(\sy H)}{(k^2-m_f^2)^3}
       \nonumber \\
      -\frac{2(\sy H)^2 k_\perp^2}{(k^2-m_f^2)^4}
     (m_f+\slsh{k_{\para}}+\slsh{k_\perp}\frac{m_f^2-k_{\para}^2}{k_\perp^2}),
\label{ferpropweak}
\eea
respectively.

There is an analogous result for gauge bosons, but since in the symmetric
phase these do not couple to the external hypermagnetic field, we do not need
to account for them.

\section{Standard Model}\label{III}

In order to consider all the contributions to the SM effective potential, we
write the Lagrangian for each sector.

The Lagrangian for the  Higgs sector is
\be
\mathcal{L}_{H}=(D_{\mu} \Phi)^{\dagger}(D^{\mu}
                \Phi)+c^2(\Phi^{\dagger} \Phi)
                -\lambda (\Phi^{\dagger} \Phi)^2,
\label{lhiggs}
\ee
where $D_{\mu}=\partial_{\mu} +ig\frac{\tau^a}{2} {
A^a_{\mu}} + i \frac{g' \sy}{2}B'_{\mu}$, $\tau^a$ are the Pauli
matrices, $B'_\mu=B_\mu+B^\ext_\mu$ and $A^a_\mu$, $B_\mu$ are the $SU(2)_L$ 
and $U(1)_Y$ gauge bosons, respectively. To allow for spontaneous symmetry
breaking the mass parameter $c$ must satisfy $c^2>0$. 

The Higgs field is a complex doublet with $\mbox{\bf \scriptsize{Y}}=+1$  
\bea
  \Phi=\frac{1}{\sqrt{2}}\left(\begin{array}{cc} 
              &\phi_3+i\phi_4  \\
              &\phi_1+i\phi_2 
           \end{array}\right),
\label{higgsdobl}
\eea
where $\phi_i$ are real scalar fields. We take $\phi_1$ as the physical Higgs
field that develops a vacuum expectation value $v$. Extremizing the tree
level potential, the parameter $c$ is related to the classical minimum
$v_{class}$
by 
\bea
   c^2=\lambda v_{class}^2.
   \label{cclass}
\eea 

For the Higgs field hypercharge conjugate doublet we use ${\tilde
\Phi}=i\sigma_2\Phi^*$.  

The kinetic energy from the $SU(2)_L$ and $U(1)_Y$  gauge bosons is
\bea
\mathcal{L}_{gb}=& & -\frac{1}{4} {\bf F}^{\mu \nu} \cdot
{\bf F}_{\mu \nu} - \frac{1}{4} F'^{\mu \nu} F'_{\mu \nu} ,
\label{kingb}
\eea
where 
\bea
   {\bf F}_{\mu \nu}&=& \partial_\mu {\bf A_\nu}- \partial_\nu {\bf
   A}_\mu-g {\bf A}_\mu \times {\bf A}_\nu \nonumber \\
   F'_{\mu \nu}&=&\partial_\mu B'_\nu- \partial_\nu B'_\mu .
\label{fieldstrength}
\eea

The Lagrangian for the fermion sector is
\bea
\mathcal{L}_{f}= & &\overline{\Psi}_R \left( i \slsh{\partial}-
                    \frac{g'}{2} \sy \slshh{B'} \right) \Psi_R + 
                    \nonumber \\
                    & &\overline{\Psi}_L \left( i \slsh{\partial} -
                    \frac{g'}{2} \sy \slshh{B'} - \frac{g}{2}{\bf \tau} 
                    \cdot
                    \bf{\slshh{A}}
                    \right) \Psi_L ,
\label{lfer}
\eea
where $\Psi_L=\frac{1}{2}(1-\gamma_5)\Psi$ is an $SU(2)$ doublet
of  left-handed fermions and $\Psi_R=\frac{1}{2}(1+\gamma_5)\Psi$ is an
$SU(2)$ singlet of right-handed fermions.

We work in the limit that all fermion masses except the top
quark mass are negligible, so that the main contribution to the Yukawa sector
is
\be
\mathcal{L}_{Yukawa}=f \overline{q_L} {\tilde \Phi} {t}_R + h.c. 
\label{lyuk}
\ee

Finally, in the $R_{\xi}$ gauge, the gauge fixing Lagrangian is
\bea
\mathcal{L}_{gf}=&-&\frac{1}{2 \xi} (\partial^{\mu}A_{\mu}^i
                            -\frac{1}{2} \xi g v \phi^i)^2- \nonumber \\
                          & &\frac{1}{2 \xi} (\partial^{\mu}B_{\mu}
                            -\frac{1}{2} \xi g' v \phi_2)^2 ,
\label{lgf}
\eea
where $i=2,3,4$ and $\xi$ is the gauge parameter. We choose to work in the
Landau gauge ($\xi = 0 $) in which the ghost fields do not acquire mass
and hence do not contribute to the $v$-dependent part of the one-loop
effective potential. Note that the effective potential is in principle a gauge
dependent object~\cite{Dolan}, however, physical quantities obtained from it
are gauge independent~\cite{Nielsen}. 

\section{Self-energies}\label{IV}

In this section we compute the SM self-energies that are in turn used for the
computation of the ring diagrams in the effective potential.  

It is well known that in the absence of an external magnetic field, the
SM thermal self-energies are gauge independent when considering only the
leading contributions in temperature~\cite{LeBellac}. However, as we will
show, when considering the effects of and external magnetic field, these
self-energies turn out to be  gauge dependent. 
\begin{figure}[t!] 
\vspace{0.4cm}
{\centering
\resizebox*{0.42\textwidth}
{0.18\textheight}{\includegraphics{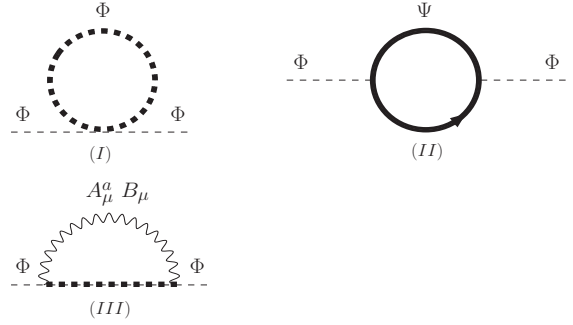}}
\par}
\caption{Self-energy Feynman diagrams for the Higgs bosons that contain
  loop particles affected by the hypermagnetic field. These particles are 
  represented by thick lines. $\Phi$ and $\Psi$ represent Higgs and Fermion
  fields whereas $A_\mu^a$ and $B_\mu$ represent the $U(1)_Y$ and $SU(2)_L$
  gauge fields, respectively.}
\label{fig1}
\end{figure}

\begin{figure}[t!] 
\vspace{0.4cm}
{\centering
\resizebox*{0.45\textwidth}
{0.10\textheight}{\includegraphics{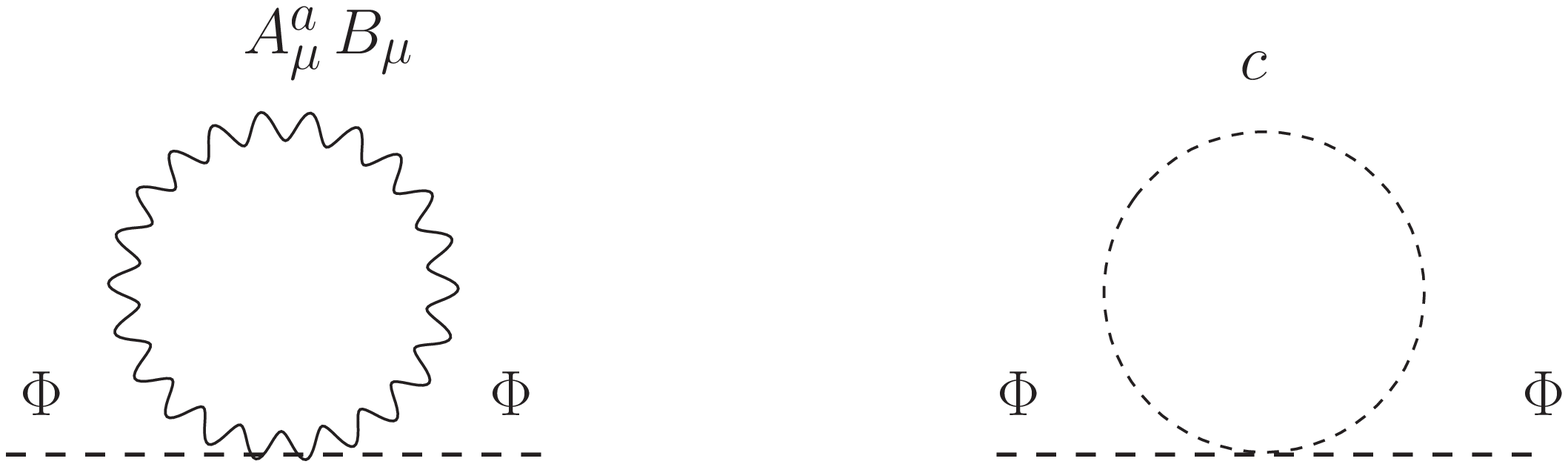}}
\par}
\caption{Self-energy Feynman diagrams for the Higgs bosons that contain
  loop particles not affected by the hypermagnetic field. These  particles are 
  represented by thin lines. $c$ represents the ghost fields. Working in the
  Landau gauge $\xi=0$, the second diagram vanishes.}  
\label{fig2}
\end{figure}

In what follows, we work in the imaginary-time formalism of thermal field
theory. First, we note that the integration over four-momenta is carried out
in Euclidean space with $k_0=ik_4$, this means that 
\bea
   \int \frac{d^4k}{(2\pi)^4} \rightarrow i \int \frac{d^4k_E}{(2\pi)^4}.
\label{minkeuc}
\eea
Next, we recall that boson energies take discrete values, namely
$k_4=\omega_n=2n \pi T$ with $n$ an integer, and thus 
\bea
 \int \frac{d^4k_E}{(2\pi)^4} \rightarrow T \sum_n \int \frac{d^3k}{(2\pi)^3}.
\eea

\subsection{Higgs boson}\label{IVa}

Figure~\ref{fig1} shows the diagrams that contribute to the Higgs boson
self-energies  affected by the hypermagnetic field. Let us explicitly compute
the momentum independent diagram shown in fig.~\ref{fig1}($I$) for a single
scalar field. In the weak field limit, its expression is 
\bea
   \Pi^{Higgs}_{(I)}&=&\frac{\lambda}{4} 
   T \sum_n \int \frac{d^3 {\bf k}}{(2\pi)^3}\nonumber\\
   &\times& 
   D_{H}(\omega_n,{\bf k};m^2\rightarrow m^2 + \Pi_1),
\label{piselfcons}
\eea
where $D_{H}$ is given by Eq.~(\ref{scalpropweak}). Notice that
Eq.~(\ref{piselfcons}) is computed self-consistently, with the
approximation that on the right hand-side,
$\Pi^{Higgs}_{(I)}\rightarrow\Pi_1$, where $\Pi_1$ is given
by~\cite{Carrington}
\bea
   \Pi_1&=& \frac{T^2}{4}
           \left\{ \frac{3}{4}g^2+\frac{1}{4}g'^2+2\lambda+f^2
           \right\},
\label{pi1}
\eea
and represents the leading temperature contribution to the scalar
self-energy. The need to compute $\Pi^{Higgs}$ self-consistently is
linked to the fact that, on the one hand, in the SM, scalar masses can vanish
as a function of $v$ and, on the other, the presence of the magnetic field
originates terms inversely proportional to these masses [see
Eq.~(\ref{autohiggsa})]. Thus, for soft momentum, where the contribution of
the ring diagrams is relevant, a naive perturbative expansion is not
sufficient. The well known correction of the infrared behavior is given by
the  plasma screening properties and in the case of Eq.~(\ref{piselfcons}) we
approximate such correction as consisting only of the leading temperature
contribution to $\Pi^{Higgs}$ which, upon resummation, take care of the
most severe infrared divergences~\cite{Dolan}. 

Using the Euclidean version of Eq.~(\ref{scalpropweak}), we have
\bea
   \Pi^{Higgs}_{(I)}&=& \frac{\lambda}{4} 
   T \sum_n \int \frac{d^3 {\bf k}}{(2\pi)^3} 
   \frac{1}{(\omega_n^2+{\bf k}^2+\tilde{m}^2)} \nonumber \\
   &\times& \left(1-\frac{(\sy H)^2}{(\omega_n^2+{\bf k}^2+\tilde{m}^2)^2} 
     \nonumber \right.\\
   &+& \left. 
   \frac{2(\sy H)^2 {\bf k}_\perp^2}{(\omega_n^2+{\bf k}^2+\tilde{m}^2)^3}
   \right),
\label{autohiggs}
\eea
where we use the short hand notation $\tilde{m}^2=m^2 + \Pi_1$.

The integrand in Eq.~(\ref{autohiggs}) contains terms whose general form
is
\bea
  I_{\alpha\beta}({\bf k},{\bf q})=\frac{1}{[\omega_n^2+{\bf
      k}^2+\tilde{m}^2]^\alpha[\omega_n^2+ 
          {({\bf k-q})}^2+\tilde{m}^2]^\beta}.\nonumber\\
\label{iab}
\eea
We make use of the Feynman parametrization to write $I_{\alpha\beta}$ as 
\bea
 I_{\alpha\beta}({\bf k},{\bf
   q})=\frac{\Gamma(\alpha+\beta)}{\Gamma(\alpha)\Gamma(\beta)} 
   \int_0^1 \frac{dx \ x^{\alpha-1} (1-x)^{\beta-1}}
   {[\omega_n^2+{\bf k'}^2(x)+{m'}^2(x)]^{\alpha+\beta}},\nonumber\\
\label{feynman}
\eea
where 
\bea
{\bf k'}(x)&=& {\bf k}-(1-x){\bf q}  \nonumber \\
{m'}^2(x)&=&\tilde{m}^2+x(1-x){\bf q}^2, 
\label{defsfeynman}
\eea
and $\Gamma$ is the Gamma function. Notice that  this parametrization is
allowed since we work in the imaginary-time formalism~\cite{weldon}.

To carry out the sum over Matsubara frequencies together with the integration
in Eq.~(\ref{autohiggs}), we perform an asymptotic expansion in the high 
temperature limit. This is done by means of a Mellin transform as
described in detail in Ref.~\cite{Bedingham}.  
The explicit result, generalized to also include the fermion case is 
\begin{widetext}
\bea
   T\sum_{n} \int \frac{d^d k}{(2\pi)^d} 
   {\bf k}^{2a}{\omega^{2t}_{n}}I_{\alpha \beta}({\bf k},{\bf q})&=&  
   \frac{(2T)(2\pi T)^{d+2a+2t-2(\alpha+\beta)} }{(4\pi)^{d/2}
   \Gamma(\alpha)\Gamma(\beta)} 
   \frac{\Gamma(\frac{d}{2}+a)}{\Gamma(\frac{d}{2})}\mu^{2\epsilon}
   \sum_{j=0}^\infty \frac{(-1)^j}{j!}
   \zeta(2(j+\alpha+\beta-t-\frac{d}{2}-a),Z)
   \nonumber \\
   &\times&  \Gamma(j+\alpha+\beta-\frac{d}{2}-a) 
   \int_0^1 dx \ x^{\alpha-1} \ (1-x)^{\beta-1}
   \left(\frac{m'(x)}{2\pi T}\right)^{2j},
\label{highT}
\eea
\end{widetext}
where $\zeta$ is the modified Riemann Zeta function, $\mu$ is the energy scale
of dimensional regularization and $d=3-2\epsilon$. For fermions the
sum runs over all integers $n$ and $Z=1/2$, while for bosons the $n=0$ term is
excluded and $Z=0$. It is important to stress that the method
advocated in Ref.~\cite{Bedingham} to perform an expression such as
Eq.~(\ref{highT}) calls for the use of dimensional regularization, which is an
appropriate method to use for non-Abelian gauge theories.   

For the terms involving the $n=0$ Matsubara frequency for bosons, we use the
result  
\begin{widetext}
\bea
 T \int \frac{d^d k}{(2\pi^d)}{\bf k}^{2a} I_{\alpha\beta}({\bf k},{\bf q})&=&
   \frac{T}{(4\pi)^{d/2}} 
   \frac{\Gamma(\frac{d}{2}+a)}{\Gamma(\frac{d}{2})}
   \frac{\Gamma(\alpha +\beta-\frac{d}{2})}{\Gamma(\alpha)\Gamma(\beta)}
   \int_0^1 dx \ x^{\alpha-1} \ (1-x)^{\beta-1} 
   \left(\frac{1}{m'(x)} \right)^{2\alpha+2\beta-d+2a}.
\label{highT0}
\eea
\end{widetext}

In terms of the functions $I_{\alpha \beta}$ defined in Eq.~(\ref{iab}) we can
write Eq.~(\ref{autohiggs})  as
\bea
   \Pi^{Higgs}_{(I)}&=& \frac{\lambda}{4} 
   T \sum_{n} \int \frac{d^3 {\bf k}}{(2\pi)^3} 
   \Big[I_{10}({\bf k},0)+I_{30}({\bf k},0)\nonumber\\
   &+&I_{40}({\bf k},0)\Big].
\label{autohiggsI}
\eea

Using Eq.~(\ref{highT}) for the terms with $n \neq 0$ and Eq.~(\ref{highT0})
for the term with $n=0$, we get
\bea
  \Pi^{Higgs}_{(I)}=\frac{\lambda}{2} T^2 
                    \left(1-\sum_{i=1}^{2} 
                    \left[\frac{3}{2\pi}\frac{\tilde{m}_i}{T}+
                    \frac{(\sy H)^2}{16 \pi T \tilde{m}_i^3}
                    \right] \right),
\label{autohiggsa}
\eea
where we have included the contribution from all scalar fields with
$\tilde{m}_i^2=m_i^2+\Pi_1$, $m_i$ standing for the scalar boson masses, given
by 
\bea
   m_1^2&=&3\lambda v^2 -c^2\nonumber\\
   m_2^2&=&m^2_3=m_4^2=\lambda v^2-c^2.
\label{higssmasses} 
\eea
In a similar fashion the contribution from the diagram in fig.~\ref{fig1}($II$)
in the {\it infrared limit}, namely $q_0=0$, $\mathbf{q} \rightarrow 0$, is
\bea
    \Pi^{Higgs}_{(II)}(0)=\frac{f^2}{4} T^2 \left( 
         1+\frac{14\zeta(3)}{(2\pi)^4} \frac{(\sy H)^2}{T^4} \right),
\label{autohiggsb}
\eea
where hereafter we use the notation ${\mathcal F}(0) \equiv {\mathcal
F}(q_0=0,{\bf q}\rightarrow 0)$ to represent the infrared limit of any
function ${\mathcal F}$. 

We point out that in Eqs.~(\ref{autohiggsa}) and~(\ref{autohiggsb}) we have
kept terms representing the leading contribution of each kind arising in the
calculation, namely, terms of order $(\sy H)^2/T^4$, $\tilde{m}_i/T$ and
$(\sy H)^2/T \tilde{m}_i^3$. For the hierarchy of scales considered, the
first kind of terms can be 
safely neglected. Recall that terms of order $\tilde{m}_i/T$ are usually
neglected in a high temperature expansion. However, since we are here
interested in keeping the leading contribution in the magnetic field strength,
we are forced to keep this kind of terms which, for a large top quark mass,
namely a large $f$, are of the same order as terms $(\sy H)^2/T
\tilde{m}_i^3$. Notice that for large values of the coupling constants, a
perturbative calculation is not entirely justified. Nevertheless, here we
consider our calculation as an analytical tool to explore this non-perturbative
domain and regard terms of order $\tilde{m}_i/T$ as an estimate of the
theoretical uncertainty of our results.

On the other hand, the diagram in fig.~\ref{fig1}($III$) is proportional to the
parameter $\xi$ and thus vanishes for our gauge parameter choice. 
Accounting also for the diagrams that are not affected by the hypermagnetic
field, and that are depicted in fig.~\ref{fig2}, the Higgs field self-energy,
in the infrared limit is given by
\bea
   \Pi^{Higgs}(0)&=&\frac{T^2}{4}
           \left\{ \frac{3}{4}g^2+\frac{1}{4}g'^2+2\lambda+f^2
           \right. \nonumber \\
           &-&
           \frac{\lambda}{\pi}\sum_{i=1}^2\left[
           \frac{3(m_i^2+\Pi_1)^{1/2}}{T}\right.\nonumber\\
           &+& \left.\left.
           \frac{(\sy H)^2}
           {8T (m_i^2+\Pi_1)^{3/2}}\right]
           \right\}.
\label{autohiggstot}
\eea

\subsection{Gauge bosons}

To express the gauge boson self-energies, 
in the presence of the
external field, we have three independent vectors to our disposal to form
tensor structures transverse to the gauge boson momentum $q^\mu$, namely
$u^\mu$, $q^\mu$ and $b^\mu$. This means that in general, these self-energies
can be written as linear combinations of nine independent
structures~\cite{Dolivo}. Since we are interested in considering 
the infrared limit, $q_0=0,{\bf q}\rightarrow0$, only $u^\mu$ and $b^\mu$
remain. Notice that the correct symmetry property for the self-energy is
$\Pi_{ab}^{\mu\nu}(q)=\Pi_{ab}^{\nu\mu}(-q)$~\cite{nievespal}. However, in the
infrared limit, this condition means that the self-energy must be symmetric
under the exchange of the Lorentz indices and therefore we can write.
\bea
 \Pi^{\mu \nu}_{ab}= \Pi^Q_{ab} Q^{\mu\nu} +
                     \Pi^R_{ab} R^{\mu\nu} +
                     \Pi^S_{ab} S^{\mu\nu} + 
                     \Pi^M_{ab} g^{\mu\nu},
\label{pibosonu1}
\eea 
where 
\bea
Q^{\mu\nu}&=&u^\mu u^\nu,\nonumber \\
R^{\mu\nu}&=&b^\mu b^\nu,\nonumber \\
S^{\mu\nu}&=&u^\mu b^\nu + u^\nu b^\mu,
\label{base}
\eea
and the transversality condition $q_\mu\Pi^{\mu\nu}_{ab}=0$ is trivially
satisfied in the infrared limit.

\begin{figure}[t!] 
\vspace{0.4cm}
{\centering
\resizebox*{0.45\textwidth}
{0.20\textheight}{\includegraphics{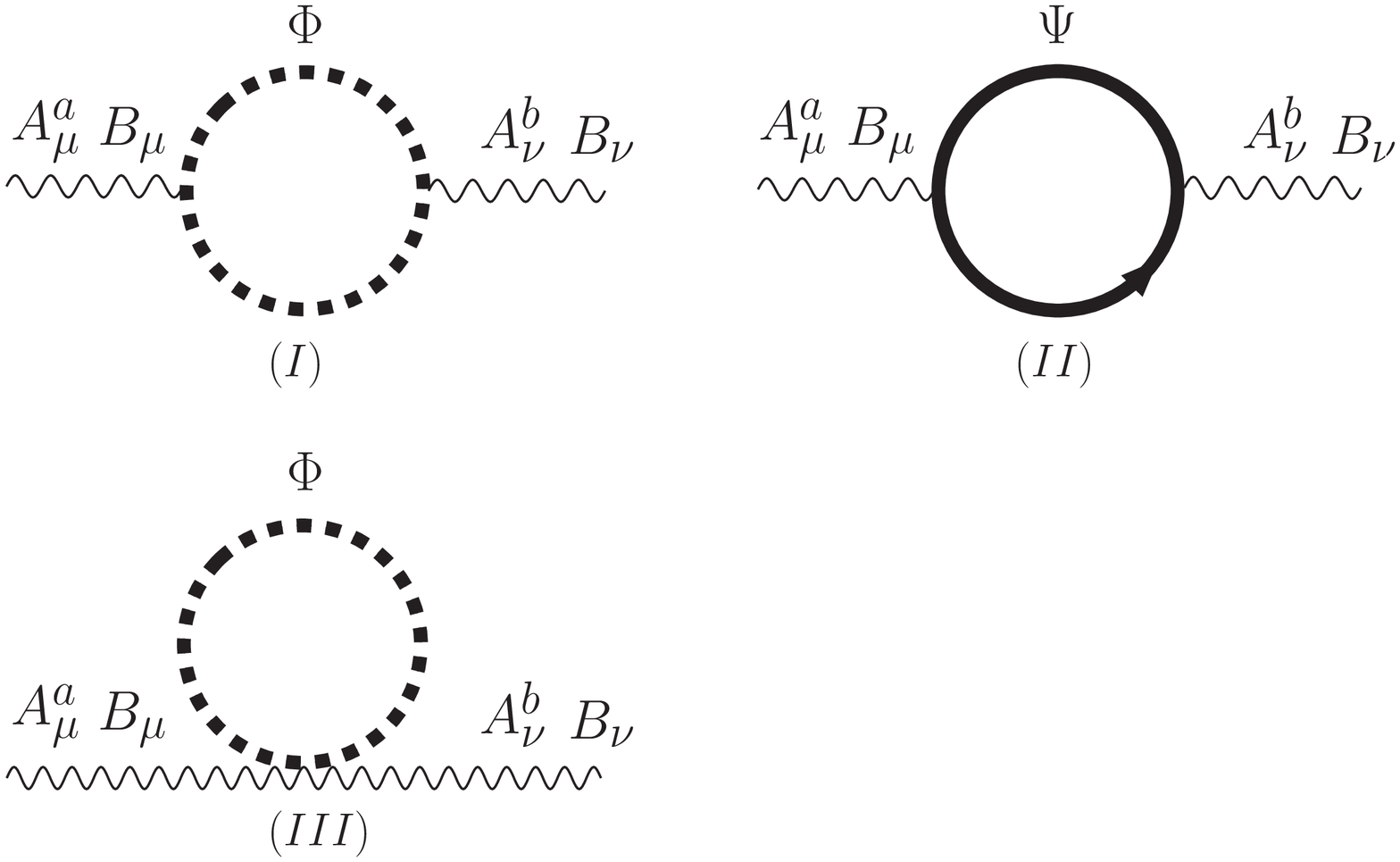}}
\par}
\caption{Self-energy Feynman diagrams for the gauge bosons that contain
  loop particles affected by the hypermagnetic field. These  particles are 
  represented by thick lines. $\Phi$ and $\Psi$ represent Higgs and Fermion
  fields whereas $A_\mu^a$ and $B_\mu$ represent the $U(1)_Y$ and $SU(2)_L$
  gauge fields, respectively.}
\label{fig3}
\end{figure}

\begin{figure}[h!] 
\vspace{0.4cm}
{\centering
\resizebox*{0.45\textwidth}
{0.20\textheight}{\includegraphics{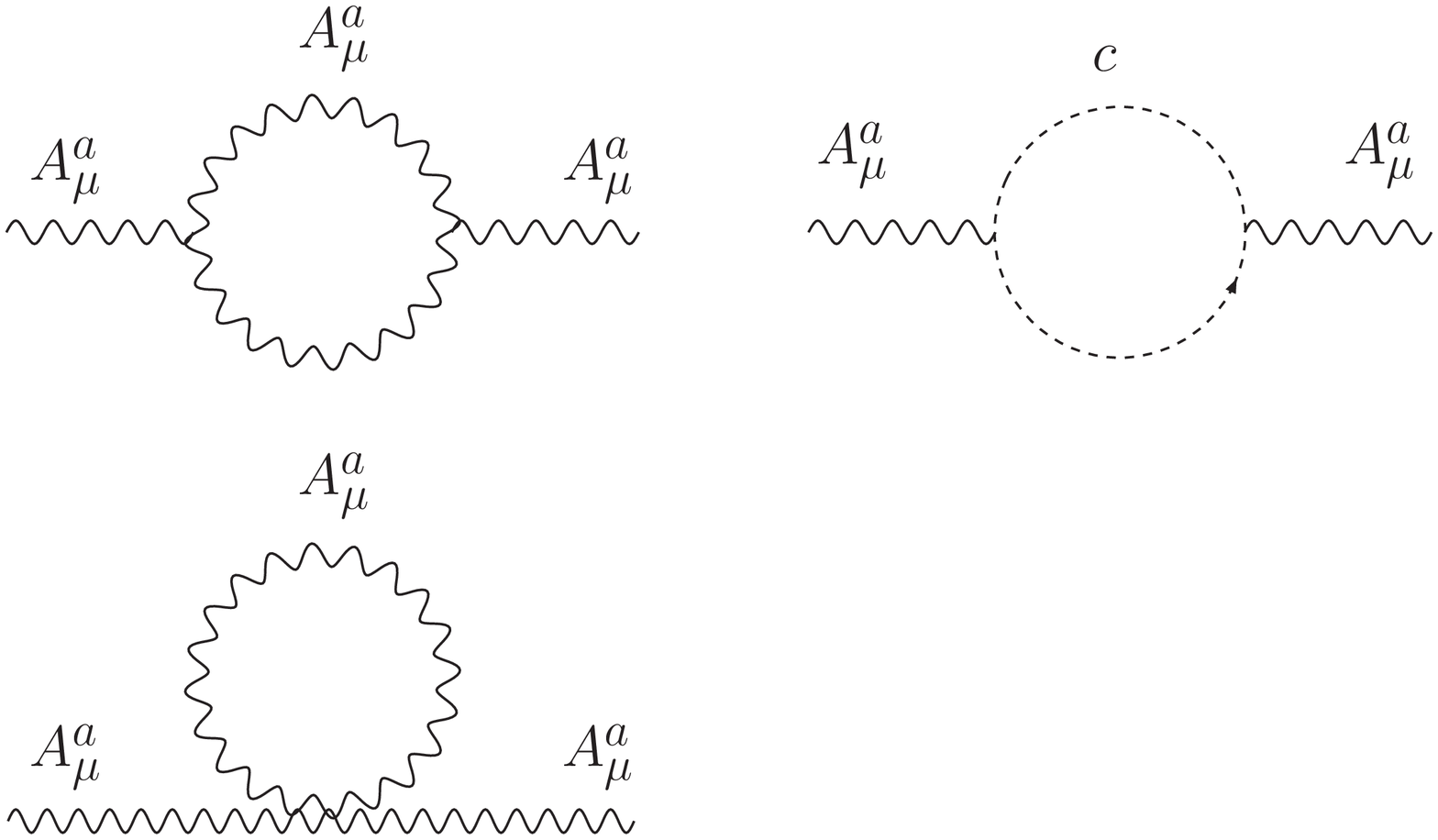}}
\par}
\caption{Self-energy Feynman diagrams for the gauge bosons that contain
  loop particles not affected by the hypermagnetic field. These particles are
  represented by thin lines. $c$ represents the ghost fields.}  
\label{fig4}
\end{figure}

Figure~\ref{fig3} shows the gauge boson self-energy diagrams that are affected
by the external hypermagnetic field. These include diagrams involving
scalars as well as fermions in the loop. Let us explicitly compute the diagram
shown in fig.~\ref{fig3}($I$) for the $B_\mu$ field. Its expression is   
\bea
   \Pi^{\mu \nu}_{(I)B}(q)&=&\left(\frac{g'}{2}\right)^2 
            \int \frac{d^4 k}{(2\pi)^4}
            (2 k^\mu - q^\mu)(2 k^\nu - q^\nu) \nonumber \\
           &\times & D_{H}(k)D_{H}(k-q).
\label{auto1a}
\eea
Notice that since the net hypercharge flowing in the loop is zero, the phase
factor in Eq.~(\ref{phase}) vanishes. 

Let us compute $\Pi^{00}_{(I)B}$ for a single scalar field with mass
$m$. Working in the rest frame of the medium,
$\Pi^{00}_{(I)B}=\Pi^{Q}_{(I)B}+\Pi^{M}_{(I)B}$. From Eq.~(\ref{auto1a}) and
at finite temperature this component is given by 
\bea
   \Pi^{00}_{(I)B}(q)&=& - g'^2 
            T \sum_n \int \frac{d^3 k}{(2\pi)^3}
            \omega_n^2 \nonumber \\
           &\times & D_{H}(\omega_n,{\bf k};m^2)
             \nonumber \\
           &\times & D_{H}(\omega_n,{\bf k-q};m^2).
\label{auto1a00}
\eea
Since the integral in Eq.~(\ref{auto1a00}) does not give rise to terms
inversely proportional to $m$, we have omitted the replacement $m^2\rightarrow
m^2+\Pi_1$.

Using the Euclidean version of the scalar propagator $D_{H}$ obtained from
Eq.~(\ref{scalpropweak}), and in terms of the functions $I_{\alpha\beta}$
defined in Eq.~(\ref{feynman}), we can write Eq.~(\ref{auto1a00}) as
\begin{widetext}
\bea
  \Pi^{00}_{(I)B}(0) = - g'^2 
        T \sum_n \int \frac{d^3 k}{(2\pi)^3} \omega_n^2 
         \left\{ I_{11}({\bf k},0)
        + (\sy H)^2\left[I_{13}({\bf k},0)
            + I_{31}({\bf k},0) 
        + 2{\bf k}^2_\perp I_{14}({\bf k},0)
           +2{\bf k}^2_\perp I_{41}({\bf k},0)\right] 
        \right\}.
\label{auto001aI}
\eea
\end{widetext}

Using Eq.~(\ref{highT}) into Eq.~(\ref{auto001aI}) and  keeping only the
leading term as discussed in Sec.~\ref{IVa}, we get 
\bea
   \Pi^{00}_{(I)B}(0) &=&  \frac{g'^2 }{12} T^2 
\label{auto001ares}
\eea
where the contribution from the four real scalar fields has been accounted
for. 

In a similar fashion, the explicit expressions for $\Pi^{00}_{(II)B}$ and
$\Pi^{00}_{(III)B}$ are computed to yield 
\bea
 \Pi^{00}_{(II)B}(0) &=& 
        \frac{5 g'^2}{3} T^2 
        \nonumber \\
\Pi^{00}_{(III)B}(0) &=& 
       \frac{g'^2}{12} T^2 \left(1 - \sum_{i=0}^{4} 
       \left[\frac{3}{4\pi}\frac{\tilde{m}_i}{T}
     + \frac{1}{32\pi}\frac{(\sy H)^2}{T \tilde{m}_i^3}\right] \right),
     \nonumber\\
\label{auto1bcres}
\eea
where in the first of the Eqs.~(\ref{auto1bcres}) we have performed the sum
over all hypercharged fermions. We stress that the origin of the terms $\sim
1/\tilde{m}_i^3$ is the topology of diagrams such as the one in
fig.~\ref{fig3}($III$) or fig.~\ref{fig1}($I$), involving a tadpole of
hypercharged scalars in the presence of the external field. In the computation
of these diagrams, we require to consider the replacement $m^2\rightarrow m^2
+ \Pi_1$ (see the discussion in Sec.~\ref{IVa}).

Adding up the above three contributions, we get
\bea
   \Pi^{00}_B=\frac{11}{6} g'^2 T^2 \left(1 -
       \sum_{i=0}^{4}\left[\frac{3}{88\pi}\frac{\tilde{m}_i}{T}
       +\frac{1}{704 \pi}\frac{(\sy H)^2}{T \tilde{m}_i^3}\right] \right).
   \nonumber\\
\label{piB}
\eea

Performing a similar exercise for the case of the ${\bf A}_\mu$ fields, for
which we also include the contribution from the diagrams that are not affected
by the hypermagnetic field, depicted in fig.~\ref{fig4}, we can
write the result for the four gauge bosons as
\bea
   {\Pi^{00}_{ab}} &=&{\tilde{g}}^2\frac{11}{6}  T^2 \left(1 - 
   \sum_{i=0}^{4}\left[\frac{3}{88\pi}\frac{(m_i^2+\Pi_1)^{1/2}}{T}\right.
   \right.\nonumber\\
   &+&\left.\left.\frac{1}{704 \pi}\frac{(\sy
   H)^2}{T(m_i^2+\Pi_1)^{3/2}}\right]\right) 
   \delta_{ab}
   \nonumber \\
   &\equiv& \tilde{g}^2\Pi_{G}(T,H) \delta_{ab}.
\label{piA}
\eea
where ${\tilde{g}}=g$ for a=b=1,2,3 and ${\tilde{g}}=g'$ for a=b=4.

As is sketched in the appendix, the other non zero components of the gauge
boson self-energy are negligible
\bea
   \Pi^{11}_{ab}=\Pi^{22}_{ab}, \ \Pi^{33}_{ab}&\sim& {\mathcal
   O}(m_i^2)\nonumber\\
   \Pi^{03}_{ab}=\Pi^{30}_{ab}&\sim& {\mathcal O}(\sy H),
\label{negligible}
\eea
which means that $\Pi^{00}_{ab}\simeq\Pi^Q_{ab}$.

Notice that in Eq.~(\ref{piA}) we have not included terms of
order ${\mathcal O}(M_{ab}/T)$, where $M_{ab}$ is the gauge boson mass
matrix. These masses are proportional to $\tilde{g}v$ whereas for large $f$,
$\tilde{m}_i$ are proportional to $fv$. Thus in this limit, terms of order
${\mathcal O}(M_{ab}/T)$ are smaller than terms of order ${\mathcal
O}(\tilde{m}_i/T)$. These terms can in principle be calculated by 
diagonalization of $M_{ab}$ or else by explicitly considering the calculation
in the basis of fields in the broken symmetry phase. We will present such
calculation in a forthcoming work along with the gauge parameter dependence of
the self-energies and effective potential. 

Although in principle the fermion self-energies are also affected by the
hypermagnetic field, as in the case of zero external field, their
contribution to the ring diagrams is subdominant in the infrared and do
not need to be taken into account.

\section{Effective Potential}\label{V}

\subsection{One-loop}

In the standard model the tree level potential is
\bea
    V_{tree}(v)= -\frac{1}{2} c^2 v^2 + \frac{1}{4} \lambda v^4.
\label{pothiggs}
\eea
To one loop,  the  effective potential (EP) receives contributions
from each sector, namely 
\bea
   V^{(1)}(v)=V^{(1)}_{gb}(v)+V^{(1)}_{Higgs}(v)+V^{(1)}_{f}(v),
\label{oneloop}
\eea
where in general each one of these contributions is given by
\bea
    V^{(1)}(v)= \frac{T}{2} \sum_n\int\frac{d^3k}{(2\pi)^3} 
               {\mbox{Tr}}\left(\ln\left[D(\omega_n,{\bf k})^{-1}\right] 
             \right),
\label{onelooptr}
\eea
with $D$ stands for either the scalar, fermion or gauge boson
propagator, and the trace is taken over all internal indices.

In the weak field limit, the contribution from the Higgs sector is given by 
\begin{widetext}
\bea
   V^{(1)}_{Higgs}&=&\sum_{i=1}^4 \frac{T}{2} 
   \sum_n\int\frac{d^3k}{(2\pi)^3}\ln
   [D_{H}^{-1}(\omega_n,{\bf k};m_i^2 \rightarrow m_i^2+\Pi_1)] 
   \simeq \sum_{i=1}^4 \frac{T}{2} \sum_n\int\frac{d^3k}{(2\pi)^3}
   \Big\{
   \ln(\omega_n^2+{\mathbf{k}}^2+m_i^2+\Pi_1)
   \nonumber \\
   &+&(\sy H)^2 \left[
   \frac{1}{(\omega_n^2+{\mathbf{k}}^2+m_i^2+\Pi_1)^2}
   -\frac{2(k_\perp^2)}{(\omega_n^2+{\mathbf{k}}^2+m_i^2+\Pi_1)^3}
   \right]\Big\}\, .
   \label{V1ap}
\eea
\end{widetext}
The first term in Eq.~(\ref{V1ap}) with $\Pi_1=0$ represents the lowest order
contribution to the effective potential at finite temperature and zero
external magnetic field, usually referred to as the boson {\it ideal gas}
contribution~\cite{LeBellac}. In order to keep track of the lowest order
corrections in $\lambda$, we set $\Pi_1=0$ in Eq.~(\ref{V1ap}). Thus for the
hierarchy of scales considered here and dropping out the zero-point energy,
this contribution is given by~\cite{Dolan}
\begin{widetext}
\bea
   \sum_{i=1}^4 \frac{T}{2}\sum_n\int\frac{d^3k}{(2\pi)^3}
   \ln(\omega_n^2+{\mathbf{k}}^2+m_i^2)
   \simeq\sum_{i=1}^4\left( -\frac{\pi^2T^4}{90}
   +\frac{m_i^2T^2}{24}-\frac{m_i^3T}{12\pi}-\frac{m_i^4}{32\pi^2}
   \ln\left(\frac{m_i}{4\pi T}\right)+{\mathcal{O}}(m_i^4)\right).
   \label{idealgas}
\eea
\end{widetext}
Notice that there are potentially dangerous terms $m_i^3$ in
Eq~(\ref{idealgas}) that can become imaginary for negative values of
$m_i$. However as we will show, these terms cancel when including the Higgs
contribution from the ring diagrams. 

The second, $H$-dependent term in Eq.~(\ref{V1ap}) vanishes
identically~\cite{nosotros}. Therefore, to one-loop order, the
contribution to the EP in the weak field case from the Higgs sector is
independent of $\sy H$ and is given by Eq.~(\ref{idealgas}).

In the weak field limit, the contribution from the fermion sector is given by
\begin{widetext}
\bea
   V^{(1)}_{f}=\sum_{i=1}^{N_f} T  \sum_n\int\frac{d^3k}{(2\pi)^3}\ln
   [S_{H}^{-1}(\omega_n,{\bf k};{m_f}_i)]
    &\simeq& \sum_{i=1}^{N_f} 2T\sum_n \int\frac{d^3k}{(2\pi)^3}
   \Big\{ \ln[\omega_n^2+{\bf k}^2+{m_f}_i^2] \nonumber \\
    &+& 
   2(\sy H)^2 \frac{\omega_n^2+k_3^2+{m_f}_i^2}
                       {(\omega_n^2+{\bf k}^2+{m_f}_i^2)^3} 
   \Big\},
\label{oneloopfer}
\eea
\end{widetext}
where the sum runs over the number of SM fermions, $N_f$, with masses
${m_f}_i=\frac{f}{\sqrt{2}}v$. We emphasize that the only fermion mass we keep
in the analysis is the top mass. 

The first term in Eq.~(\ref{oneloopfer}) represents the fermion {\it ideal gas}
contribution~\cite{LeBellac}, which is explicitly given by~\cite{Dolan}
\begin{widetext}
\bea
   \sum_{i=1}^{N_f} 2 T \sum_n\int\frac{d^3k}{(2\pi)^3}
   \ln[\omega_n^2+{\bf k}^2+{m_f}_i^2]  
   &\simeq& \sum_{i=1}^{N_f} \Big\{-7\frac{\pi^2 T^4}{180} 
   +\frac{{m_f}_i^2T^2}{12} +\frac{{m_f}_i^4}{16 \pi^2} 
   \ln\left(\frac{{m_f}_i^2}{T^2}\right)
   +{\mathcal{O}}({m_f}_i^4) \Big\}.
\label{idealgasfer}
\eea
\end{widetext}
The second term in Eq.~(\ref{oneloopfer}) is subdominant, after taking care of 
renormalization and running of the coupling constant, as we sketch in the
Appendix. Therefore, to one-loop order, the
contribution to the EP in the weak field case from the fermion sector is
only  given by Eq.~(\ref{idealgasfer}).

Finally, since in the symmetric phase the gauge bosons do not couple to the
external field, their contribution to the EP is given by~\cite{Carrington}
\bea
   V^{(1)}_{gb}&=&\sum_{G}\frac{T}{2}\sum_n\int\frac{d^3k}{(2\pi)^3}
   {\mbox{Tr}} \ln[{(D^{\mu\nu}_{ab})}^{-1}(\omega_n,{\bf k};m_G)]\nonumber \\
   &\simeq& \frac{T}{2} \sum_n\int\frac{d^3k}{(2\pi)^3}
   \Big\{ 6 \ln(\omega_n^2+{\mathbf{k}}^2+m_W^2)  \nonumber \\
   &+& \
    3 \ln(\omega_n^2+{\mathbf{k}}^2+m_Z^2) \nonumber \\
    &+& 
    2 \ln(\omega_n^2+{\mathbf{k}}^2)
      \Big\},
\label{idealgasgb}
\eea
where on the right-hand side in the first line, the index $G$ runs over the
four SM gauge bosons and thereafter, we have used the mass eigenbasis with
$m_W^2=g^2v^2/4 $ and $m_Z^2=(g^2+g'^2)v^2/4$. The
factors in front of each contribution correspond to the two $W'$s, the $Z$ and
the photon polarizations.

Using Eq.~(\ref{idealgas}) into Eq.~(\ref{idealgasgb}), the contribution to
the EP from the gauge boson sector is
\begin{widetext}
\bea
   V^{(1)}_{gb}=
   -11 \frac{\pi^2T^4}{90}
   +3\frac{(2m_W^2+m_Z^2)T^2}{24}-\frac{(2m_W^3+m_Z^3)T}{12\pi}
   -6\frac{m_W^4}{32\pi^2}\ln\left(\frac{m_W}{4\pi T}\right)
   -3\frac{m_Z^4}{32\pi^2}\ln\left(\frac{m_Z}{4\pi T}\right)
   +{\mathcal{O}}(m_W^4).
\label{idealgasgbT}
\eea
\end{widetext}

In our expressions for the 1-loop potential, ultraviolet divergences are
absorbed by the standard renormalization procedure~\cite{arnold}.

\subsection{Ring diagrams}

\begin{figure}[t!] 
\vspace{0.4cm}{\centering
\resizebox*{0.46\textwidth}
{0.08\textheight}{\includegraphics{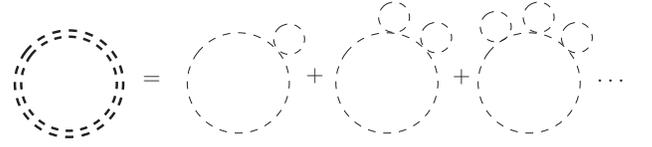}}
\par}
\caption{Schematic representation of ring diagrams, that consist in the
  resummation of successive insertions of self-energies in vacuum bubbles.}
\label{fig5}
\end{figure}

It is well known that the next order correction to the EP
comes from the so called {\it ring diagrams}. These are schematically depicted
in fig.~\ref{fig5}. Their contribution to the EP is given by 
\bea
   V^{(ring)}_{Higgs}(v)&=&-\sum_{i=1}^4 \frac{T}{2}
               \sum_n\int\frac{d^3k}{(2\pi)^3} \mbox{Tr}
               \nonumber \\
              &\times&  \left\{\sum_{N=1}^{\infty}
               \frac{1}{N}\left[-D_{H}(\omega_n,{\bf k};m_i)\Pi^{Higgs}(0)
               \right]^N \right\} \nonumber \\
              &=&-\sum_{i=1}^4 \frac{T}{2} \sum_n\int\frac{d^3k}{(2\pi)^3} 
               \nonumber \\
             &\times&  
             \mbox{Tr} \ \ln[1+\Pi^{Higgs}(0)D_{H}(\omega_n,{\bf k};m_i)],
\label{ringhiggs}
\eea
for the scalar case, and by 
\bea
   V^{(ring)}_{gb}(v)&=&-\frac{T}{2}\sum_n\int\frac{d^3k}{(2\pi)^3}
               \nonumber \\
               &\times& \mbox{Tr} \left\{\sum_{N=1}^{\infty}
               \frac{1}{N}\left[-\Pi^{ab}_{\mu\lambda}(0)
               D_{bc}^{\lambda\nu}(\omega_n,{\bf k})
               \right]^N \right\},\nonumber \\
\label{ringgb}
\eea
for the gauge boson case. The dominant contribution comes from the mode $n=0$
in Eqs.~(\ref{ringhiggs}) and~(\ref{ringgb})~\cite{LeBellac}. The scalar case
has been treated in detail in Ref.~\cite{nosotros} for the weak field limit
and here we just quote the result
\begin{widetext}
\bea
   V^{(ring)}_{Higgs}\!\!=\!\!- \sum_{i=1}^4 \left\{ 
           \frac{T}{12\pi}\left[\left(m_i^2+
           \Pi_1\right)^{3/2} - m_i^3 \right] 
         - \frac{(\sy H)^2}{4\pi}\left(\frac{\Pi_1}{48}\right)
           \left(\frac{T}{(m_i^2+\Pi_1)^{3/2}}\right) \right\},
\label{VringBno0}
\eea
\end{widetext}
where the contributions from all scalars has been accounted for.

In order to compute the contribution from the gauge bosons, we have to
diagonalize the matrix product
$\Pi^{ab}_{\mu\lambda}(0)D_{bc}^{\lambda\nu}(\omega_n,{\bf k})$ in 
Eq.~(\ref{ringgb}). In the gauge group space, we notice that since
$\Pi^{ab}(0)$ is, according to Eq.~(\ref{piA}), diagonal we can
use the same matrix that diagonalizes the mass matrix, given explicitly for
instance in Ref.~\cite{arnold}. On the other hand in Lorentz space, the
Euclidean version of the gauge boson propagator in the Landau gauge can be
written as 
\bea
    D_{\mu\nu}&=& \Big\{P^L_{\mu\nu}+P^T_{\mu\nu}\Big\} \frac{1}{k^2+m^2},
\label{propbas}
\eea
where 
\bea
    P^T_{00}=P^T_{0i}=0&  &P^T_{ij}=\delta_{ij}-{\bf \hat{k}}_i{\bf \hat{k}}_j 
    \nonumber \\
    P^L_{\mu\nu}=\delta^{\mu\nu} \!\!&-&\!\!
             \frac{k_\mu k_\nu}{k^2}-P^T_{\mu\nu}.
\label{translong}
\eea
It is easy to see that considering only the dominant term in the product
$\Pi_{\mu\lambda}(0)D^{\lambda\nu}(\omega_n,{\bf k})$ one gets
\bea
   \Pi_{\mu_\lambda} D^{\lambda\nu}&=& \frac{\Pi^{Q}_{ab}}{k^2+m^2}
              \left[1+\frac{(k\cdot u)^2}{k^2}\right]Q_\mu^\nu.
\label{propbas2}
\eea
Considering the $n=0$ term and taking the trace, Eq.~(\ref{ringgb}) gives 
\begin{widetext}
\bea
    V^{(ring)}_{gb}(v) &=&-\sum_{G} \frac{1}{2} T \int\frac{d^3k}{(2\pi)^3} 
                          \ln[1+\frac{\Pi^{Q}_{ab}}{{\bf k}^2+m_G^2}] 
                      \nonumber \\ 
                       &=&-\frac{T}{12\pi}\left\{
                  2m_W^3(T,H)+m_Z^3(T,H)+m_{\gamma}^3(T,H)
                  - 2m_W^3-m_Z^3 \right\},
\label{ringgb2}
\eea
\end{widetext}
where $m_G(T,H)$ are explicitly given by 
\begin{widetext}
\bea
   m_W^2(T,H)&=&m_W^2+g^2\Pi_G \nonumber \\
   m_Z^2(T,H)&=&\frac{1}{2}\left\{ m_Z^2+(g^2+g'^2)\Pi_G
           + \sqrt{[m_Z^2+(g^2+g'^2)\Pi_G]^2-
           8\frac{g^2g'^2\Pi_G}{(g^2+g'^2)}
            [m_Z^2+\frac{(g^2+g'^2)}{2}\Pi_G]}\right\} 
              \nonumber \\
   m_\gamma^2(T,H)&=&\frac{1}{2}\left\{ m_Z^2+(g^2+g'^2)\Pi_G
           - \sqrt{[m_Z^2+(g^2+g'^2)\Pi_G]^2-
           8\frac{g^2g'^2\Pi_G}{(g^2+g'^2)}
            [m_Z^2+\frac{(g^2+g'^2)}{2}\Pi_G]}\right\}. \nonumber \\
\label{massT}
\eea
\end{widetext}
Note that both the temperature as well as the magnetic field corrections to
the gauge boson masses are encoded in $\Pi_G$. When $H \rightarrow 0$
the gauge boson masses tend to their usual thermal masses~\cite{Carrington}
\bea
     m_W^2(T,H\rightarrow 0)     &\rightarrow& 
     m_W^2(T)\nonumber \\
     m_Z^2(T,H\rightarrow 0)     &\rightarrow& m_Z^2(T)  \nonumber \\
     m_\gamma^2(T,H\rightarrow 0)&\rightarrow& m_\gamma^2(T).
\label{massesT0}
\eea
where $m_G^2(T)$ can be obtained from Eqs.~(\ref{piA}) and~(\ref{massT}) by
setting $H=0$.  

To bring about the explicit dependence of $V_{gb}^{(ring)}$ on the
hypermagnetic field strength, let us expand the the cube of the
gauge boson masses appearing in Eq.~(\ref{ringgb2}) up to order $(\sy
H)^2$, which is our working order. From the expressions in 
Eqs.~(\ref{massT}) this gives

\begin{widetext}
\bea
  V^{(ring)}_{gb}(v) &=& -\frac{T}{12\pi}\left\{
                  2m_W^3(T)+m_Z^3(T)+m_{\gamma}^3(T)
                  - 2m_W^3-m_Z^3  \right\} \nonumber \\
        &+&\frac{T}{12\pi}\Big\{  
        \frac{3}{4}
        \left[2 g^2m_W(T)+(g^2+g'^2)(m_Z(T)R_Z-m_\gamma(T)R_\gamma)\right] 
        \nonumber \\
        &\times& \frac{T}{384 \pi} \sum_{i=1}^4   
        \left(\frac{1}{\left(m_i^2+\Pi_1\right)^{3/2}}
        \right)\Big\} (\sy H)^2
\label{massesTexpanded}
\eea
\end{widetext}
where 
\bea
   R_Z&=& \frac{2m_Z^2(T)}{m_Z^2(T)-m_\gamma^2(T)}-
      \frac{g^2g'^2}{(g^2+g'^2)^2}
      \frac{m_Z^2(T)+m_\gamma^2(T)}{m_Z^2(T)-m_\gamma^2(T)} \nonumber \\
   R_\gamma&=& \frac{2m_\gamma^2(T)}{m_Z^2(T)-m_\gamma^2(T)}-
      \frac{g^2g'^2}{(g^2+g'^2)^2}
      \frac{m_Z^2(T)+m_\gamma^2(T)}{m_Z^2(T)-m_\gamma^2(T)}.  \nonumber \\  
\label{Rs}
\eea

The final expression for the effective potential is obtained by adding up the
results in Eqs.~(\ref{pothiggs}), (\ref{idealgas}), (\ref{idealgasfer}),
(\ref{idealgasgbT}), (\ref{VringBno0}) and~(\ref{massesTexpanded}).
\bea
   V(v)&=&V_{tree}(v)+V^{(1)}_{Higgs}+V_f^{(1)}+V^{(1)}_{gb} \nonumber \\
         &+& V^{(ring)}_{Higgs}+V^{(ring)}_{gb}      
\label{finalpot}
\eea
Notice that the dangerous terms $m_i^3$ which come from $V^{(1)}_{Higgs}$
cancel with the corresponding terms coming from $V^{(ring)}_{Higgs}$. The
cubic term in the masses of the gauge bosons do not cancel but since these
masses are positive definite, these terms do not pose any problem. Also, in
order for the terms involving the square of the bosons' thermal mass to be
real, the temperature must be such that
\bea
   T>T_1&\equiv&\sqrt{\frac{-16 m_1^2(v=0)}{3g^2+g'^2+8\lambda+4f^2}},
   \label{bound1}
\eea
which defines a lower bound for the temperature. A more restrictive bound is
obtained by requiring that $\sy H < \tilde{m}^2$ for the weak field expansion
to work. This condition translates into the bound
\bea
   T>T_2&\equiv&\sqrt{\frac{\sy H-16 m_1^2(v=0)}{3g^2+g'^2+8\lambda+4f^2}}.
\label{bound2}
\eea 

The relevant factor that enhances the order of transition, present both in
$V^{(ring)}_{Higgs}$ and $V^{(ring)}_{gb}$, is $(\sy H)^2/\tilde{m}_i^3$
which can be traced back to the boson self-energy diagrams involving a tadpole
of hypercharged scalars in the presence of the external field.

\section{Symmetry Breaking}\label{VI}

In order to quantitatively check the effect of the magnetic field during the
EWPT, we proceed to plot $V_{eff}$ as a function of the vacuum expectation
value $v$. For the analysis we use $g'=0.344$ and $g=0.637$, $m_Z=91$ GeV,
$m_W=80$ GeV, $f=1$, $\lambda=0.11$ which corresponds to the current bound on
the Higgs mass. 

\begin{figure}[t!] 
\vspace{0.4cm}
{\centering
\resizebox*{0.4\textwidth}
{0.20\textheight}{\includegraphics[angle=0]{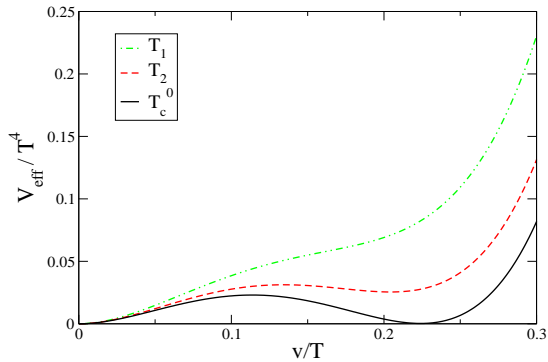}}
\par}
\caption{(Color on line) $V_{eff}$ as a function of $v$ for different
temperatures and 
$H =0$.  At high temperature ($T_1$) the symmetry is restored.
Decreasing the temperature ($T_2$) causes $V_{eff}$ to develop a secondary
minimum that becomes degenerate with the original one for a critical
temperature $T_c^0<T_1<T_2$, where $v=v^0$.  Lowering further the temperature
below $T_c^0$ produces the system to spinodaly decompose. For the analysis we
use $g'=0.344$ and $g=0.637$, $m_Z=91$ GeV, $m_W=80$ GeV, $f=1$,
$\lambda=0.11$.} 
\label{fig6}
\end{figure}

Figure~\ref{fig6} shows $V_{eff}$ for different temperatures and
$H =0$. This figure shows  the usual behavior whereby at high
temperature the symmetry is restored and, when decreasing the temperature,
$V_{eff}$ develops a secondary minimum that becomes degenerate with the
original one for a critical temperature $T_c^0$, where $v=v^0$.  Lowering
further the temperature below $T_c^0$ produces the system to spinodaly
decompose.

\begin{figure}[t!] 
\vspace{0.4cm}
{\centering
\resizebox*{0.4\textwidth}
{0.20\textheight}{\includegraphics[angle=0]{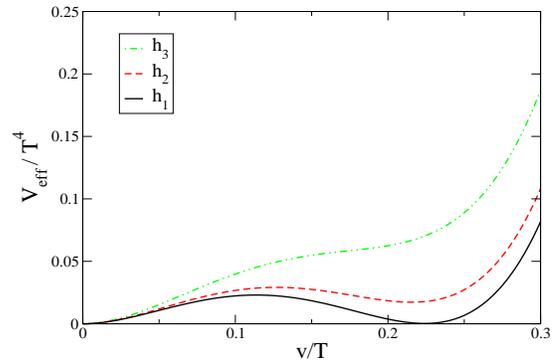}}
\par}
\caption{(Color on line) $V_{eff}$ as a function of $v$ for constant $T$ and
  different values
  $H=h\,(100\ \mbox{GeV})^2$ where $h_1=0$, $h_2=0.03$ and $h_3=0.06$.
  The choosen temperature $T$ is the critical temperature for $h_1=0$. For
  higher field strengths, the phase transition is delayed, favoring higher
  values of $\langle v\rangle/T_c$. For the analysis we use $g'=0.344$ and
  $g=0.637$, $m_Z=91$ GeV, $m_W=80$ GeV, $f=1$, $\lambda=0.11$.}
\label{fig7}
\end{figure}

To study the effect of the hypermagnetic field on the effective potential, we
parametrize $H=h\,(100\ \mbox{GeV})^2$. In what follows, we always choose
values for $h$ that comply with the condition $h(\sy \,(100\
\mbox{GeV})^2)<\tilde{m}_1^2(v=0)$, as required from our hierarchy of scales,
since the smallest possible mass is always that of a scalar boson in the
symmetric phase.  Figure~\ref{fig7} shows $V_{eff}$ for different field
strengths $H_3>H_2>H_1$ and constant $T$ where $T$ is taken as the critical
temperature for the $h=0$ case. We can see that for higher field strengths,
the phase transition is delayed, favoring higher values of the ratio of the
vacuum expectation value $\langle v\rangle$ in the broken symmetry phase to
the critical temperature $T_c$. This makes the transition stronger first order
as the strength of the field increases.

\begin{figure}[t!] 
\vspace{0.4cm}
{\centering
\resizebox*{0.4\textwidth}
{0.20\textheight}{\includegraphics[angle=0]{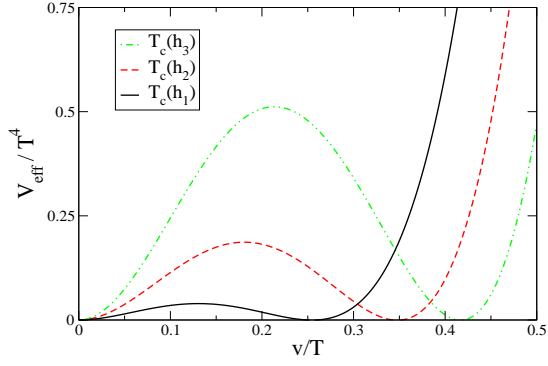}}
\par}
\caption{(Color on line) $V_{eff}$ as a function of $v$ for different
hypermagnetic field 
strengths $h_3>h_2>h_1$ at their corresponding critical
temperatures 
$T_c(h_3)<T_c(h_2)<T_c(h_1)$, where $H=h\,(100\ \mbox{GeV})^2$. We
note that increasing the intensity of the field, the barrier 
between minima becomes higher and that the ratio $\langle v\rangle /T$ becomes
larger. For the analysis we use $g'=0.344$ and
$g=0.637$,  $m_Z=91$ GeV, $m_W=80$ GeV, $f=1$, $\lambda=0.11$.}
\label{fig8}
\end{figure}

Figure~\ref{fig8} shows a comparison of $V_{eff}$ for different field
strengths $H_3>H_2>H_1$ at their corresponding critical
temperatures with $T_c(h_3)<T_c(h_2)<T_c(h_1)$ for $h_3=0.25$, $h_2=0.2$ and
$h_1=0.1$. From this
figure we also note that increasing the intensity of the field, the barrier
between minima becomes higher and the ratio $\langle v\rangle /T$ becomes
larger, making the EWPT become stronger first order~\cite{Petropoulos}. 

The effect of the hypermagnetic field is therefore twofold: first it delays
the beginning of the phase transition and second the Higgs vacuum expectation
value in the broken symmetry phase becomes larger. Both effects favor the
suppression of the sphaleron transition rate and therefore help the freezing
of a possible baryon asymmetry  during the EWPT.   

\section{Discussion and Conclusions}\label{VII}

In conclusion, we have shown that in the presence of an external magnetic
field, the EWPT becomes stronger first order. Our treatment has been
implemented for the case of weak fields for the hierarchy of scales $\sy H
\ll m^2 \ll T^2$ where $m$ is taken as a generic mass involved in the
calculation. Notice that when this relation is applied to the case of the
scalar masses, $m^2$ should be regarded as $\tilde{m}^2$. We have explicitly
worked with the degrees of freedom in the 
symmetric phase where the external magnetic field belongs to the $U(1)_Y$
group and therefore properly receives the name of {\it hypermagnetic}
field. The calculation is carried out up to the contribution of ring
diagrams to the effective potential at finite temperature. To include the
effects of this external field, we have made use of the Schwinger proper-time
method. In this way, the contribution from all Landau levels has been
accounted for. We have carried out a systematic expansion up to order $(\sy
H)^2$. The presence of the external hypermagnetic field gives rise to terms
in the effective potential proportional to $1/\tilde{m}_i^3$, where
$\tilde{m}_i^2=m_i^2+\Pi_1$, coming from tadpole diagrams in the boson
self-energies where the loop particle is a hypercharged scalar and $m_i$ are
their masses. These terms are the relevant ones for the strengthening of the
order of the phase transition since $\tilde{m}_i(v)$ is small for small values
of $v$, which enhances the curvature of the effective potential near $v=0$. 

The presence of the hypermagnetic field has two simultaneous effects: first it
delays the beginning of the phase transition as compared to the case with no
hypermagnetic field. Second, the Higgs vacuum expectation value in the broken
symmetry phase becomes larger. Both effects favor the suppression of the
sphaleron transition rate and therefore help the freezing  of a possible
baryon asymmetry during the EWPT. Although our 
study suggests that
for reasonable values of the magnetic field strength the ratio $\langle
v\rangle/T$ does not reach the necessary values to avoid the sphaleron
erasure, we should keep in mind that we have worked under very restrictive
assumptions, derived from our assumed hierarchy of scales. Lifting some of the
restrictions, in particular the weakness of the magnetic field compared to
the mass scale, could open a window to make the above ratio larger, while
remaining within the observational bounds. This is a subject under current
study and will be reported elsewhere. 
The calculation has been carried out in the framework of perturbation theory
but in order to make contact with the values for the masses of the top quark
and the current bounds on the Higgs mass, in the numerical analysis we have
used large values for the coupling constants, in particular we have considered
the top Yukawa coupling $f=1$. In this sense, our calculation has to be
considered as an analytical tool to explore this non perturbative regime.

Our results go along the same direction as the ones obtained at
tree~\cite{Giovannini} and one-loop~\cite{Elmfors}levels. We should however
point out that the same problem has been also treated in 
Refs.~\cite{Skalozub1,Skalozub2} in the context of strong external
magnetic/hypermagnetic fields. These authors conclude that the magnetic field
gives rise to logarithmic terms of the ratio of the temperature to the fermion
masses and 
that for light fermions, these terms increase the {\it inertia} of the Higgs
field producing the phase transition to become second order. Similar
logarithmic terms appear in a weak field expansion of the effective potential
in Ref.~\cite{Persson} although, as pointed out by the authors of that work,
their weak field expansion is not necessarily reliable since it is given in
terms of a Borel summable rather than a convergent series. In this work we
have not come across such terms but a detailed comparison between the methods
used in the above mentioned works with ours to include the effects of the
magnetic field is certainly called for. Also the study of the gauge parameter
dependence and the inclusion of small but not necessarily negligible terms
proportional to the gauge boson masses is a natural extension of this
work. Progress in this direction is being made and we will report on it
elsewhere~\cite{Jorge}.

\section*{Acknowledgments}

We acknowledge useful conversations with M.E. Tejeda, S. Sahu A. Raya and
J.L. Navarro during the genesis and realization of the work. Support has been
received in part by DGAPA-UNAM under PAPIIT Grant No. IN107105 and by
CONACyT-M\'exico under Grant No. 40025-F.

\section*{Appendix}

First we sketch the computation of the diagram depicted in
fig.~\ref{fig3}($II$) representing the fermion contribution to the one-loop
gauge boson self-energy tensor up to order $(\sy H)^2$, where it is not
evident that the off-diagonal components are negligible. We use the notation
$^n\Pi_{\mu\nu}$, $n=0,1,2$, where $n$ is the power of $\sy H$. To zeroth
order we have
\bea
   ^0\Pi_{\mu\nu}&=&\left(\frac{g}{4}\right)^2
           T\sum_{n}\int \frac{d^3k}{(2\pi)^3}\nonumber\\
           &\times&\frac{16 k_\mu k_\nu-8k^2\delta_{\mu\nu}}{[k^2+{m_f}_i^2]
           [(k-p)^2+{m_f}_i^2]}.
\label{A1}
\eea
Using the result in Eq.~(\ref{highT}) for fermions 
we get 
\bea
    ^0\Pi_{\mu\nu}&=&\left(\frac{g}{4}\right)^2
    \Big\{\left(\frac{2}{3}T^2 + \frac{{m_f}_i^2}{2\pi^2}
        \left[ 
        \frac{1}{\epsilon}+\gamma_E+\ln{\frac{4\mu^2}{\pi T^2}}-1 
        \right]\right) \nonumber \\
        &\times&\delta_{\mu 0}\delta_{\nu 0}
        +\frac{{m_f}_i^2}{(4\pi)^2} 
        \left[\frac{2}{3}+\frac{1}{\epsilon}+\gamma_E+
        \ln{\frac{4\mu^2}{\pi T^2}}\right]
        \delta_{i\mu}\delta_{j\nu}
        \Big\}\!.\nonumber\\
\label{A1a}
\eea
To first order we have
\bea
   ^1\Pi_{\mu\nu}&=& -8(\sy H) \left(\frac{g}{4}\right)^2
         T\sum_{n}\int \frac{d^3k}{(2\pi)^3}\nonumber\\
         &\times&
         \left[ k_3(k_\mu u_\nu+k_\nu u_\mu) + 
         \omega_n(k_\mu b_\nu+k_\nu b_\mu)\right.\nonumber\\ 
         &-& \left. 2\omega_n k_3\delta_{\mu\nu}    
         \right][I_{12}({\bf k},{\bf q})+I_{21}({\bf k},{\bf q})]\nonumber\\
         &+&8i\epsilon_{\mu\nu\alpha\beta}\left[k_3 k^\alpha u^\beta
         +\omega_n k^\alpha b^\beta\right]\nonumber\\
         &\times&[I_{12}({\bf k},{\bf q})-I_{21}({\bf k},{\bf q})].
\label{A2}
\eea
Using again Eq.~(\ref{highT}) 
we explicitly get 
\bea
   ^1\Pi_{\mu\nu}&=&\left(\frac{g}{4}\right)^2 \frac{(\sy H)}{2\pi^2}
      \left[
       \frac{1}{\epsilon}+\gamma_E+\ln{\frac{4 \mu}{\pi T}}+\frac{1}{3} 
      \right] \nonumber \\
      &\times& (\delta_{\mu 3}\delta_{\nu 0} +\delta_{\mu 0}\delta_{\nu 3}).
\label{A2a}
\eea
To second order we get contributions from two terms
\bea
   ^2\Pi_{\mu\nu}^{a}&=& 8(\sy H)^2 \left(\frac{g}{4}\right)^2
         T\sum_{n}\int \frac{d^3k}{(2\pi)^3}
         \left[k_3^2(2u_\mu u_\nu-\delta_{\mu\nu}) \right. \nonumber \\
         &+& \left. 2\omega_n k_3(u_\mu b_\nu+u_\nu b_\mu)
         +\omega_n^2(2b_\mu b_\nu-\delta_{\mu\nu})\right]\nonumber \\ 
         &\times& I_{22}({\bf k},{\bf q}),
\label{A3}
\eea
and
\bea
    ^2\Pi_{\mu\nu}^{b}&=&-8(\sy H)^2 \left(\frac{g}{4}\right)^2
         T\sum_{n}\int \frac{d^3k}{(2\pi)^3} \nonumber \\
     &\times& \left[
     4k_\mu k_\nu(k_3^2-\omega_n^2) -2((k_\mu b_\nu +k_\nu b_\mu)k_3
     \right. \nonumber \\
     &-& \left. 
     (k_\mu b_\nu +k_\nu b_\mu)\omega_n)(\omega_n^2+{\bf k}^2)\right]
     \nonumber \\
     &\times& \left(I_{41}({\bf k},{\bf q})+I_{14}({\bf k},{\bf q})
     \right) \nonumber \\
     &-& 2 i\epsilon_{\mu\nu\alpha\beta}\left[k_3 b^\alpha k^\beta
     -\omega_n u^\alpha k^\beta\right]k^2 \nonumber \\
     &\times&  (I_{41}({\bf k},{\bf q})-I_{14}({\bf k},{\bf q})).
\label{A4}
\eea
Using once more Eq.~(\ref{highT}) 
we get
\bea
   ^2\Pi_{\mu\nu}^{a}&=& -8\left(\frac{\sy H}{T}\right)^2
             \left(\frac{g}{4}\right)^2
             \frac{7 \xi(3)}{48\pi^4}
             \left\{ \delta_{\mu 0}\delta_{\nu 0} \right. \nonumber \\
             &+&\left. (2 
              + 3b^i b^j ) \delta_{\mu i} \delta_{\nu j}
             \right\},
\label{A3a}
\eea
and
\bea
    ^2\Pi_{\mu\nu}^{b}&=&8\left(\frac{\sy H}{T}\right)^2 
             \left(\frac{g}{4}\right)^2
             \frac{7 \xi(3)}{144\pi^4}
             \left\{ \delta_{\mu i}\delta_{\nu j} \right. \nonumber \\ 
             &+& \left. 6\delta_{\mu 3}\delta_{\nu 3}
            -9\delta_{\mu 0}\delta_{\nu 0} \right\}.
\label{A4a}   
\eea

From Eq.~(\ref{A1a}), we see that the leading contribution corresponds to the
term proportional to $T^2$ and thus the result in the first of
Eqs.~(\ref{auto1bcres}), after summing over all hypercharged fermions. After
factorizing the leading contribution, we also
notice that the terms proportional to $({m_f}_i/T)^2$ come together with a
factor reminiscent of renormalization and running of the fermion mass and
couplings with the scale $\mu$. We do not carry out this procedure in detail
but when absorbing these factors into the redefinition of the fermion mass, we
see that corrections to the leading term are proportional to $({m_f}_i/T)^2$
and therefore for the values taken by $v$ during the EWPT, these corrections
can be safely neglected. 

From Eq.~(\ref{A2a}), and by the same procedure as for the case of the zeroth
order terms, the off-diagonal terms contribute an amount proportional to $\sy
H/T^2$, compared to the leading term and thus for the hierarchy of scales
considered, this term can also be neglected.

Finally from Eqs.~(\ref{A3a}) and~(\ref{A4a}) we can see that the second order
contributions are proportional to  $(\sy H/T^2)^2$ and therefore can also
be safely neglected.

Next, we sketch the computation of the hypermagnetic field contribution to the
fermion one-loop effective potential $V_f^{(1)}$. According to
Eq.~(\ref{oneloopfer}), this contribution is of second order in $\sy H$.
\bea
 ^2V_f^{(1)}=2(\sy H)^2 T \sum_n \int \frac{d^3k}{(2\pi)^3}
       \frac{\omega_n^2+k_3^2+{m_f}_i^2}
       {(\omega_n^2+{\bf k}^2+{m_f}_i^2)^3}.
\label{A5}
\eea
Using Eq.~(\ref{highT}) 
we get
\bea
  ^2V_f^{(1)}&=&\frac{2(\sy H)^2}{32 \pi^2}
  \left[\frac{1}{\epsilon}-\frac{7\xi(3)}{2\pi^2}\frac{{m_f}_i^2}{T^2}
  \right.\nonumber\\
  &+&\left.\ln(\frac{4 \mu^2}{\pi T^2})+\gamma_E+\frac{2}{3} 
  \right].
\label{A6}
\eea
After renormalization and running of the couplings with the scale $\mu$ we see
that this contribution is of order $(\sy H/T^2)^2$ and can also be safely
neglected.


\begin{thebibliography}{99}

\bibitem{Sakharov} 
A.D. Sakharov, Pis'ma Zh. \'Eksp. Teor. Fiz. 5, 32 (1967) [JETP Lett. 5, 24
(1967)].

\bibitem{Gavela} 
M.B. Gavela, P. Hern\'andez, J. Orloff and O. P\`ene, Mod. Phys. Lett. A {\bf
9}, 795 (1994). 

\bibitem{Kajantie1}
K. Kajantie, M. Laine, K. Rummukainen and M. Shaposhnikov, Nucl. Phys. {\bf
B466}, 189 (1996).

\bibitem{Klinkhamer}
F.R. Klinkhamer and N.S. Manton, \prd {\bf 30}, 2212 (1984).

\bibitem{reviewsEWPT}
For comprehensive reviews on EW baryogenesis see for example:
M. Trodden, Rev. Mod. Phys. {\bf 71}, 1463 (1999);
A. Megevand, Int. J. Mod. Phys. D {\bf 9}, 733 (2000).

\bibitem{Petropoulos}
N. Petropoulos, ``Baryogenesis at the electroweak phase transition'',
hep-ph/0304275.

\bibitem{31}
For comprehensive reviews see P.P. Kronberg, Rep. Prog. Phys. {\bf 57}, 325
(1994); R. Beck, A. Brandedenburg, D. Moss, A. Shukurov and D. Sokoloff,
Annu. Rev. Astron. Astrophysics, {\bf 34}, 155 (1996); C.L. Carilli and
G.B. Taylor, Ann. Rev. Astron. Astrophysics, {\bf 40}, 319 (2002).


\bibitem{Barrow}
J.D. Barrow, P. Ferreira and J. Silk, \prl {\bf 78}, 3610 (1997).

\bibitem{Yamazaki}
D.G. Yamazaki, K. Ichiki, T. Kajino and G.J. Mathews, {\it  Constraints on the
evolution of the primordial magnetic field from the small scale cmb angular
anisotropy}, astro-ph/0602224. 

\bibitem{Barrow2}
See J.D. Barrow, R. Maartens and C.G. Tsagas, {\it Cosmology with inhomogeneous
magnetic fields}, astro-ph/0611537 and references therein.

\bibitem{Grasso}
D. Grasso and H.R. Rubinstein, Phys. Lett. {\bf 379} 73, (1996)

\bibitem{elreview}
See for example: G. Piccinelli and A. Ayala, Lect. Notes Phys. {\bf 646},
293-308 (2004).

\bibitem{Giovannini}
M. Giovannini and M. E. Shaposhnikov, \prd {\bf 57}, 2186 (1998).

\bibitem{Elmfors}
P. Elmfors, K. Enqvist and K. Kainulainen, Phys. Lett. B {\bf 440},
269 (1998). 

\bibitem{Kajantie}
K. Kajantie, M. Laine, J. Peisa, K. Rummukainen and M. Shaposhnikov,
Nucl. Phys. B {\bf 544}, 357 (1999).

\bibitem{Pallares} 
A. Ayala, J. Besprosvany, G. Pallares and G. Piccinelli, \prd {\bf 64},
123529 (2001); A. Ayala, G. Piccinelli and G. Pallares, \prd {\bf66},
103503 (2002), A. Ayala and J. Besprosvany, Nucl. Phys. {\bf B651},
211 (2004). 

\bibitem{Dario}
D. Comelli, D. Grasso, M. Pietroni and A. Riotto, Phys. Lett. {\bf B458}, 304
(1999). 

\bibitem{Skalozub1}
V. Skalozub and V. Demchik, {\it  Electroweak phase transition in strong
magnetic fields in the Standard Model of elementary particles};
hep-th/9912071. 

\bibitem{Skalozub2}
V. Skalozub and M. Bordag, Int. J. Mod. Phys. A {\bf 15}, 349
(2000). 

\bibitem{Carrington} M.E. Carrington, \prd {\bf 45}, 2933 (1992).

\bibitem{Schwinger} J. Schwinger, Phys. Rev. {\bf 82}, 664 (1951). 

\bibitem{nosotros}
A. Ayala, A. S\'anchez, G. Piccinelli and S. Sahu,
\prd {\bf 71}, 023004 (2005). 

\bibitem{Tzuu}
T.-K. Chyi, C.-W. Hwang, W.F. Kao, G.L. Lin, K.-W. Ng and J.-J. Tseng,
Phys. Rev. D {\bf 62}, 105014, (2000).

\bibitem{Maartens}
R. Maartens, Pramana {\bf 55}, 575 (2000).

\bibitem{Ambjorn} 
J. Ambj{\o}rn and P. Olesen, Nucl. Phys. {\bf B315}, 606 (1989).

\bibitem{Dolan} L. Dolan and R. Jackiw, \prd {\bf 9}, 3320 (1974).

\bibitem{Nielsen} N. K. Nielsen, Nucl. Phys. {\bf B} 101, 173 (1975).

\bibitem{LeBellac} M. Le Bellac {\it Thermal Field Theory},
Cambridge University Press (1996).

\bibitem{weldon} H.A. Weldon, \prd {\bf 47}, 594 (1993).

\bibitem{Bedingham} D.J. Bedingham, \lq\lq Dimensional regularization
and Mellin summation in high temperature calculations'',
hep-ph/0011012.

\bibitem{Dolivo}
J.C. D'Olivo, J.F. Nieves, S. Sahu, Phys. Rev. D {\bf 67}, 025018
(2003). 

\bibitem{nievespal} J.F. Nieves and P.B. Pal, \prd {\bf 39}, 652 (1989).

\bibitem{arnold}  P. Arnold and O. Espinosa \prd {\bf 47}, 3546 (1993).

\bibitem{Persson} P. Elmfors, D. Persson and B.-S. Skagerstam,
 Astropart. Phys. {\bf 2} 299 (1994).

\bibitem{Jorge}
J.L. Navarro, A. Ayala, G. Piccinelli, A. S\'anchez and M.E. Tejeda, in
progress. 

\end{thebibliography}
\end{document}